\input harvmac

\def\and{ {\rm and} }

\def\be{\begin{equation}}
\def\ba{\begin{eqnarray}}
\def\ea{\end{eqnarray}}   
\def\gr{\nabla}
\def\ee{\end{equation}}
\def\to{\rightarrow}
\def\tmu{$TH\epsilon\mu\;$}
\def\pd{\partial}
\def\nn{\not\!}
\def\sp{\overline\psi}
\def\api{{\alpha\over \pi}}
\def\al{\alpha}

%-------------------
% title page
%-------------------
%
\Title{\vbox{\baselineskip12pt
\hbox{WATPHYS TH-96/07}
%\hbox{gr-qc/96?????}
}}
{\vbox{\centerline{Testing the Equivalence Principle in
the Quantum Regime} }}
\baselineskip=12pt
\centerline{
Catalina Alvarez\footnote{$^1$}{Internet:
calvarez@avatar.uwaterloo.ca}}
\bigskip
\centerline {and}
\bigskip
\centerline {Robert Mann\footnote{$^2$}{Internet:
mann@avatar.uwaterloo.ca}}
\medskip
\centerline{\sl Department of Physics}
\centerline{\sl University of Waterloo}
\centerline{\sl Waterloo, Ontario N2L 3G1 Canada}

\bigskip
\centerline{\bf Abstract}
\medskip

We consider possible tests of the Einstein Equivalence Principle
for physical systems in which quantum-mechanical vacuum energies
cannot be neglected. Specific tests include a
search for the manifestation of non-metric effects in Lamb-shift
transitions of Hydrogenic atoms and in anomalous magnetic moments of
massive leptons. We discuss how current experiments already set
bounds on the violation of the equivalence principle in this sector
and how new (high-precision) measurements 
of these quantities could provide  further 
information to this end.

%\draftmode
\Date{}

\gdef \jnl#1, #2, #3, 1#4#5#6{ {\it #1~}{\bf #2} (1#4#5#6) #3}

\lref\gpb{J.P. Blaser {\it et.al.}, {\sl STEP Assessment Study Report},
ESA Document SCI (94)5, May (1994)}
%{\sl Gravity Probe B}, Stanford University Preprint, ed. C.W.F. Everitt. 
\lref\Gb{R.S. van Dyck, P.B. Schwinberg and H.G. Dehmelt,
Phys. Rev. Lett. {\bf 59}, 26 (1987).}
\lref\Bailey{J. Bailey {\it et.al.}, Phys. Lett. {\bf B68}, 191 (1977);
J. Bailey {\it et.al.}, Nuovo Cimento {\bf A9}, 369 (1972).}
\lref\Kino{T. Kinoshita, {\sl Quantum Electrodynamics} 
(World Scientific, Singapore,1990).}
\lref\anti{ M.H. Holzscheiter, T. Goldman, M.M. Nieto, 
LA-UR-95-2776, (1995), hep-ph/9509336}
\lref\catmag{C. Alvarez and R.B. Mann,  WATPHYS-TH95/10,  gr-qc/9511028}
\lref\utpal{R.B. Mann and U. Sarkar, Phys. Rev. Lett. {\bf 76}
865 (1996); M. Gasperini, Phys. Rev. {\bf D39}, 3606 (1989)}
\lref\jody{M. Gabriel, M. Haugan, R.B. Mann and J. Palmer, 
Phys. Rev. Lett. {\bf 67} 2123 (1991)}
\lref\kll{K.S. Thorne, A. P. Lightman and D. L. Lee, Phys. 
Rev. D {\bf 7}, 3563 (1973).}
\lref\redshift{R. F. C. Vessot and M.W. Levine, Gen. Relativ. Gravit. {\bf 10},
181 (1979).}
\lref\PLC{J. D. Prestage et al., Phys. Rev. Lett. {\bf 54},  2387 (1985);
 S. K. Lamoreaux et al.,  {\it ibid.} {\bf 57}, 3125 (1986);
T. E. Chupp et al., {\it ibid.} {\bf 63}, 1541 (1989).}
\lref\will{C. M. Will, {\it Theory and Experiment in Gravitational Physics},
2nd edition (Cambridge University Press, Cambridge, 1992).}
\lref\HW{M. P. Haugan and C. M. Will, Phys. Rev. Lett. {\bf 37}, 1 (1976);
Phys. Rev D. {\bf 15}, 2711 (1977).}
\lref\catlamb{C. Alvarez and R.B. Mann, WATPHYS-TH95/02, gr-qc/9507040}
\lref\thmu{A. P. Lightman and D. L. Lee, Phys. Rev. D {\bf 8}, 364 (1973).}
\lref\gabriel{ M. D. Gabriel and M. P. Haugan, Phys. Rev. D {\bf 41}, 
2943 (1990).}
\lref\Hughes{R.J. Hughes, Contemporary Physics {\bf 34} 177 (1993).}
\lref\Eides{M.I. Eides and V.A. Shelyuto, Phys. Rev. {\bf A52} 954 
(1995).}
\lref\Schiff{L. Schiff, Proc. Nat. Acad. Sci. {\bf 45} 69 (1959).}

Our understanding of gravitation is built upon the foundations of
the Equivalence Principle.  Originally regarded as the cornerstone
of mechanics by Newton, and used later by Einstein in the development
of general relativity, it has come to be understood as the basis for
the notion that spacetime has a unique operational geometry. It 
consequently ensures that the effects of gravity on matter can be 
described in a purely geometric fashion.

Of the several existing variants of the equivalence principle, it is
the Einstein Equivalence Principle (EEP) which plays a pivotal role in
this regard.  This principle has three components as follows. 
The first is that all freely falling
bodies ({\it i.e.} bodies which are not acted upon by non-gravitational
forces such as electromagnetism and which are small enough so that tidal
effects are negligible) move independently of their composition
(the Weak Equivalence Principle, or WEP).  The second component is the 
assertion that the results of any non-gravitational experiment (such as the measurement
of an electromagnetic current in a wire) are independent of where and
when in the universe it is carried out (Local Position Invariance, or 
LPI), and the third component is the
assertion that such results are independent of the velocity of the
freely falling reference frame in which the experiment is performed
(Local Lorentz Invariance (LLI)).  
Metric theories (such as general relativity and Brans-Dicke Theory) 
realize the EEP by endowing spacetime with a 
symmetric, second-rank tensor field $g_{\mu\nu}$ that couples universally 
to all non-gravitational fields {\kll}, so that that in a local
freely falling frame the three postulates are satisfied. By
definition, non-metric theories do not have this feature:  
they violate universality and so permit observers performing local
experiments to detect effects due to their position and/or velocity
in an external gravitational environment.

Each of the three components of EEP have been  subjected to severe
experimental scrutiny. Empirical limits on WEP violation are typically
set by torsion balance experiments, whereas limits on LPI and LLI
violation are set by gravitational red-shift \redshift\ and
atomic physics experiments {\PLC}\ respectively, 
all to varying degrees of precision. The universality
of gravitational redshift has been verified to 1 part in 5000 
{\redshift}, WEP to 1 part in $10^{12}$ {\will} and LLI to
1 part in $10^{21}$ {\PLC}. Significantly improved 
levels of precision are anticipated in future experiments \gpb .

Impressive as these limits are, the dominant form of mass-energy 
governing the systems these experiments study is nuclear electrostatic 
energy, although violations of the EEP due to other forms of energy 
(virtually all of which are associated with baryonic matter) 
have also been estimated {\HW}. However there are many physical systems 
dominated by other forms of mass  energy
for which the validity of the equivalence principle has yet to be 
empirically checked, including matter/antimatter systems {\anti}, 
(hypothesized) dark matter, photons of differing polarization {\jody}, 
massive leptons, neutrinos {\utpal},
second and third generation matter,
and quantum vacuum fluctuations. Comparatively little is known 
about empirical limits on EEP-violation in these other sectors 
{\Hughes}.

We describe in this paper the results of an approach for examining 
potential violations of the EEP due to effects which are peculiarly 
quantum-mechanical in origin ({\it i.e.} are due solely to 
radiative corrections). Effects of this type include
Lamb-shift transition energies in Hydrogenic atoms and 
anomalous magnetic moments of massive leptons. Tests of the EEP 
in this sector push the confrontation between quantum mechanics and
gravity ever closer, providing us with qualitatively new empirical 
windows on the foundations of gravitational theory.

The action appropriate for Quantum Electrodynamics in
a background gravitational field (GQED) is
\eqn\act{
\!\!S\!=\!\int\!\! d^4x\sqrt{-g}\!\left[ \sp(i\nn\gr\!+\!e\nn\! A\!-\!m)\psi
\!-\!{1\over 4}F_{\mu\nu}F^{\mu\nu}\right]}
where 
$\gr$ is the covariant derivative, 
$F_{\mu\nu} \equiv A_{\nu,\mu} - A_{\mu,\nu}$ and
${\nn\! A}= e^a_\mu \gamma_a A^\mu$, $e^a_\mu$ being the tetrad 
associated with the metric. Our approach is to extend this action
to the wide class of non-metric theories described by the
\tmu formalism {\thmu}. This formalism (which has as its limiting
case all metric theories) assumes that the external gravitational
environment of a given physical system is desribed by a static, spherically
symmetric metric which does not necessarily couple universally to all
forms of matter. More concretely,
$$
g_{\mu\nu}= {\rm diag}(-T,H,H,H) \quad {\rm and} \quad
F_{\mu\nu}F^{\mu\nu} = 2(\epsilon E^2-B^2/\mu)
$$
where $\vec E\equiv-\vec\gr A_0-\pd\vec A/\pd t$ and
$\vec B \equiv \vec\gr\times\vec A$. 
$\epsilon$ and $\mu$ are
arbitrary functions of the Newtonian background potential
$U= GM/r$ (which approaches unity as $U\to 0$) as are $T$ and $H$,
which in general will depend upon the species of particles within the
system, which we shall take to be massive leptons.

The action \act\ will in general depend upon the velocity 
$\vec u$ of a given (sub)atomic system relative to the preferred frame
(whose coordinates define the form of \act )
as well as the \tmu parameters.  This dependence can be obtained
using a Lorentz transformation to transform fields and coordinates from 
the preferred frame to the rest frame of the (sub)atomic system, whose
small spacetime size permits us to ignore spatial variations in the
\tmu parameters.  Analysis shows that, upon local rescaling of physical
parameters, it is only the electromagnetic sector of
the action that depends explictly on $\vec{u}$ and the 
dimensionless parameter $\xi_\ell \equiv 1-c_\ell^2 
= 1-H_0/T_0\epsilon_0\mu_0 $, ``0'' denoting evaluation at the system's 
center of mass, with $c_\ell$ the ratio of the limiting speed of 
lepton `$\ell$' to the speed of light \catlamb . 
It is straightforward to check that all divergences in GQED can be 
renormalized by proper redefinitions of the parameters of the theory, 
(which now include the \tmu functions) and that the Ward identities 
are satisfied. This is a consequence of gauge invariance.
In extracting predictions from GQED, we note that
the natural scale for $\xi$ is set by the magnitude of $U$, which 
empirically is much smaller than unity, permitting 
a perturbative analysis in $\xi$.

The Lamb shift is an energy shift caused by quantum vacuum fluctuations
between normally degenerate states in a Hydrogenic atom. 
To extract the prediction for this effect in GQED, it is necessary to
solve the field equations for the electromagnetic vector potential
produced by a pointlike nucleus of charge $Ze$ at rest in the moving frame.
Employing previously established techniques {\gabriel}\ yields the
result that this degeneracy is lifted before radiative corrections 
are introduced \catlamb .  This non-metric energy shift is isotropic in 
$\vec u$ and vanishes when $\vec{u} = 0$.  Evaluating the relevant
radiative corrections to the required accuracy 
$O(\xi_\ell)O(\vec u^2)O(\al(Z\al)^4 )$ entails a very lengthy and tedious 
calculation, leading to an expression for a gravitationally modified
Lamb shift energy $\Delta E_L(\xi_\ell,\vec{u})$ which is no longer
isotropic in $\vec{u}$.

Upper bounds on $\xi_\ell$ from current experiments can be obtained by
assuming that EEP-violating contributions to $\Delta E_L$ are
bounded by the current level of precision for the Lamb shift 
{\Eides}. Using the accepted upper limit $|\vec{u}| < 10^{-3}$ for
the preferred frame velocity {\will}\ we find \catlamb\ the 
dominant non-metric contribution to $\Delta E_L$ is due to purely 
radiative corrections, yielding the bound $|\xi_e| < 10^{-5}$. 
If we assume that positrons and
electrons do not have equivalent couplings to the gravitational field
{\Schiff}, we find that there is an additional radiative contribution 
to $\Delta E_L$ due to vacuum polarization. Making the same comparisons 
as above, we find the most stringent bound on this quantity to be 
$|\xi_{e^+} | < 10^{-3}$ from present Lamb shift experiments.  

In the case of anomalous magnetic moments, we find that
an evaluation of the Feynman amplitude related to the elastic 
scattering of a lepton by a static external field yields an effective 
interaction term in the GQED Hamiltonian 
\eqn\inter{
\!\! H_\sigma\!\!=\!\!-{e\over 2m}\!\{g\vec S\cdot\vec B+
g_*\vec S\cdot\vec u\,\vec B\cdot\vec u\}
\!\!\equiv\!\! -\Gamma^{ij}\!S_iB_j}
with
$g\equiv 2+\api[1+\xi_\ell(1+ {\gamma^2 \over 6}(1+7\vec u^2))]$, 
$g_*\equiv-\api\xi_\ell{4\over 3}\gamma^2$, and
$\gamma^2 = (1-|\vec{u}|^2)^{-1}$,
where $\vec S\equiv {\vec\sigma\over 2}$ is the spin operator.
The presence of preferred frame effects induces a new type of tensorial 
coupling between the magnetic field and the spin described by 
$\Gamma_{ij}$.  

From this we find that the precession frequency 
of the longitudinal spin polarization 
$\vec{S}\cdot\vec{\beta}$ is $\simeq {eB\over m} a$, where
\eqn\aa{
\!a\!=\!{\al\over 2\pi}\!\{1+\xi [1+{\gamma^2\over 6}\!(\!1+7(V^2\!+\beta^2)
\!-8V^2\!\cos^2\Theta)\}}
and where $\vec{\beta}$ is the lepton velocity with respect to the laboratory
system which moves in the preferred frame with velocity $\vec{V}$,
whose angle with the magnetic field is $\Theta$.
Assuming the  EEP-violating contributions 
to  $a$ are bounded by the current level of precision 
for gyromagnetic anomalies (and that $|\vec{V}| < 10^{-3}$) 
then the discrepancy between the best empirical 
and theoretical values for the electron yields  the bounds \catmag\
$|\xi_{e^{-}}| <  10^{-8}$  and 
$|\xi_{e^{-}} - \xi_{e^+}| < 10^{-9}$,
the latter following from a
comparison of positron and electron magnetic moments.
For muons, a similar analysis yields $|\xi_{\mu^{-}}| < 10^{-8}$
and $|\xi_{\mu^{-}} - \xi_{\mu^+}| < 10^{-8}$.
To our knowledge these limits are the most stringent yet noted
for these parameters.

Non-metric effects also induce oscillations in 
the spin polarization component ($S_B$) parallel to $B$. The
ratio between the temporal average of this quantity
and the initial polarization of the beam can be estimated to be
\catmag\ 
$\delta_\ell = {\langle S_B\rangle \over S}
\sim\xi_\ell V\beta\cos\Theta\gamma^2$.
In highly relativistic situations this effect is enhanced, and can be
estimated by considering a 
typical experiment with $V\sim 10^{-3}$, where for
electrons ($\beta\sim 0.5$), and so $\delta_e\sim 10^{-11}$;
and for muons ($\beta=0.9994$), $\delta_\mu\sim 10^{-8}$.
In both cases the corresponding present constraints for
$\xi_\ell$ were employed. Improved measurements of this quantity
for different values of $\Theta$ should afford the opportunity of
putting tighter constrains on the $\xi_\ell$ parameters.
The rotation of the Earth will have the effect of converting
this orientation dependence into a time-dependence
with a period related to that of the sidereal day. 

Additional  empirical information can also be extracted from 
$\Delta E_L$ and $\Gamma_{ij}$ by evaluating their associated 
gravitational redshift parameters. Analysis \catlamb\ shows that
these parameters are two linearly independent combinations of
$\Gamma_0\!\equiv\! {T_0\over T_0'}\ln[{T\epsilon^2 \over H}]'\vert_0$
and $\Lambda_0\!\equiv\!{T_0\over T_0'}\ln[{T\mu^2\over H}]'\vert_0$.
Redshift experiments can therefore set bounds on independent regions
of $(\Gamma_0,\Lambda_0)$ parameter space in the lepton sector.
However this will be a challenge to experimentalists because of the 
small redshift due to earth's gravity ($< 10^{-9}$) 
and the intrinsic uncertainties of excited states of Hydrogenic atoms. 
One would at least need to perform these experiments in a
stronger gravitational field (such as on a satellite in close solar 
orbit) with 1-2 orders-of-magnitude improvement in precision.

To summarize, violation of gravitational universality 
modifies radiative corrections to lepton--photon interactions
in a rather complicated way, giving rise to several novel effects.
Refined measurements of atomic vacuum transitions and
anomalous magnetic moments can provide
an interesting new arena for investigating the validity of the EEP
in physical regimes where quantum field theory cannot be neglected.
It will be a challenge to set new empirical bounds on such effects 
in the next generation of experiments.

\bigskip
\noindent{Acknowledgments}
\medskip
This work was supported in part 
by the Natural Sciences and Engineering Research
Council of Canada.  We are grateful to C.W.F. Everitt, M. Haugan, 
E. Hessels and  C.M. Will for interesting discussions at various 
stages of this work.

\listrefs
\end